\documentclass[prb,superscriptaddress,twocolumn,amsmath,amssymb]{revtex4}
\usepackage{graphicx}
\usepackage{dcolumn}
\usepackage{bm}
\usepackage{color} % for nice colors in the links
\definecolor{darkblue}{rgb}{0,0,0.5}
\definecolor{lila}{rgb}{0.3,0,0.3}
\definecolor{turq}{rgb}{0,0.1,0.4}
\usepackage{url} % that hyperlinks may be hyphenated correctly
\begin{document}

\title{Quantum logic for control and manipulation of molecular ions using a frequency comb}
\author{S. Ding}
\affiliation{Centre for Quantum Technologies, National University of Singapore, 3 Science Drive 2, Singapore 117543}
\author{D. N. Matsukevich}
\email{phymd@nus.edu.sg}
\affiliation{Centre for Quantum Technologies, National University of Singapore, 3 Science Drive 2, Singapore 117543}
\affiliation{Department of Physics, National University of Singapore, 2 Science Drive 3, Singapore 117542}
\date{\today}
\begin{abstract}
Due to their rich level structure, molecules are well-suited for probing time variation of fundamental constants \cite{junye1,junye3,kozlov2}, precisely measuring parity violation \cite{willeke,kozlov3} and time-reversal non-invariance effects \cite{cornell}, studying quantum mechanical aspects of chemical reactions \cite{softley}, and implementing scalable quantum information processing architectures \cite{andre}. Molecular ions are particularly attractive for these applications due to their long storage times and the near-perfect isolation from environment that result in long coherence times required to achieve high measurement precision and reduce systematic errors.  However, the control of molecular quantum states remains a challenge. Based on quantum logic techniques \cite{schmidt,rosenband},  we propose a scheme for preparation, manipulation, and detection of quantum states of single molecular ions. The scheme relies on coherent coupling between internal and motional degrees of freedom of the molecular ion via a frequency comb laser field, while detection and cooling of the motion of ions is done via a co-trapped atomic ion.
\end{abstract}
%\pacs{37.10.Mn, 33.20.-t, 82.37.Vb}
\maketitle

Traditional methods of state detection for neutral atoms and atomic ions are difficult to apply in the case of molecules. It is hard to find a nearly closed cycling transition, which limits the applicability of standard fluorescence detection methods. Other detection methods such as resonance-enhanced multiphoton dissociation \cite{rempd} are destructive in nature. In addition, a large number of available energy levels  also complicates the preparation of a molecule in a single quantum state. While the translational motion of  molecular ions can be sympathetically cooled to milikelvin temperatures, the rotational and vibrational degrees of freedom will still be in equilibrium with the environment. Since typical spacing between rotational levels of a molecule is on the order of 10-100 GHz, much lower than the thermal energy $k T / h\simeq 6$ THz at room temperature, several molecular states are populated in the equilibrium with the environment. Due to the interaction of a molecule with black body radiation, internal degrees of freedom reach the equilibrium on the timescale of the order of minutes \cite{drewsen,schiller,vogelius}.

Several methods of preparing molecular ions in a single quantum state have recently been demonstrated, including optical pumping assisted by black body radiation \cite{drewsen,schiller}, and sympathetic cooling of a translational degree of freedom for molecular ions formed in a particular rovibrational state \cite{willitsch}. Other promising techniques such as optical pumping using broadband light \cite{viteau,odom}, sympathetic cooling of molecular ions in a cloud of ultracold neutral atoms \cite{hudson}, and cavity cooling \cite{morigi,lev} have been proposed and are currently under active experimental investigation.

Some proposals \cite{lazarou,vogelius,vogelius2} also consider coherent coupling between the internal states of molecular ions and their motion, and subsequent cooling of the motional degree of freedom as a way of preparing and detecting the molecular states. However, coherent coupling between an internal quantum state of the molecule and the motion of the ions in these schemes is typically achieved using pulses produced by CW lasers. This approach may be difficult to implement experimentally given the number of molecular states that one needs to address.  Recently, coherent coupling between hyperfine states of atomic Yb$^{+}$ ion and its motion using a frequency comb laser field has been demonstrated \cite{hayes}. Due to the large bandwidth and precise control afforded by the frequency comb, it is particularly suitable for addressing of multiple molecular energy levels \cite{junye4}. 

\begin{figure}
\centering
\includegraphics[width=8cm]{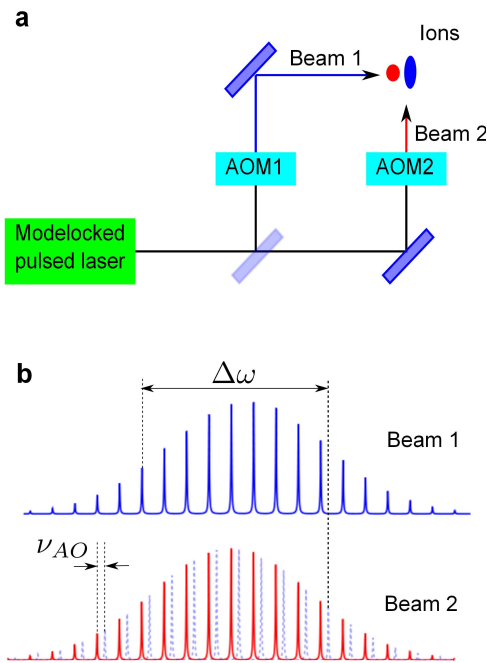}
\caption{The proposed experimental setup. a) Light emitted by a modelocked pulsed laser with the repetition rate $f_{rep}$ is split into two beams, sent through the acousto-optical modulators to offset relative  frequency between two beams by $\nu_{AO}$ and is incident upon the atomic and molecular ions from two different directions. b) Spectrum of the frequency combs. If resonance conditions are satisfied the setup can drive a stimulated Raman transition between a pair of energy levels $\Delta \omega$ apart.}
\label{fig:setup}
\end{figure}

Here we propose an experimental scheme to prepare, manipulate and detect the internal states of a cold trapped molecular ion based on quantum logic techniques \cite{rosenband,schmidt,damop}. The experimental setup that we consider is shown in Fig. \ref{fig:setup}. We confine a diatomic molecular ion together with an atomic ion in a standard rf-Paul trap. Due to Coulomb interaction, atomic and molecular ions share common modes of motion. The motion of the molecular and atomic ions are cooled to the ground state by a sideband cooling\cite{monroe}.  Two laser beams generated by a modelocked pulsed laser with a repetition rate $f_{rep}$, are offset by frequency $\nu_{AO}$ with acousto-optical modulators, and are focused on the ions from two different directions.

In order to drive a stimulated Raman transition between two quantum states separated by $\Delta \omega$, the repetition rate $f_{rep}$ and offset frequency $\nu_{AO}$ should satisfy resonance condition
\begin{equation}
\Delta \omega = M f_{rep} \pm \nu_{AO}.
\end{equation}
Here integer number $M$ is the comb index. The time averaged resonance Rabi frequency between two states coupled by an off-resonance pulsed laser is\cite{hayes}
\begin{equation}
\Omega = \Omega_0 \left( \frac{ \Delta \omega \tau}{e^{\Delta \omega \tau / 2} - e^{- \Delta \omega \tau / 2} } \right),
\label{eq:omega}
\end{equation}
where $\tau$ is the pulse duration,  $\Omega_0 = s \gamma^2 / 2 \Delta$, and $s = \bar I / I_{sat}$, $\bar I$ is the average intensity,  $I_{sat}$ is the saturation intensity, $\Delta$ is the detuning of the pulsed laser from an excited state, and $\gamma$ is the excited state decay rate. The Rabi frequency is suppressed if the energy splitting between two states $\Delta \omega$ is greater than the bandwidth of the pulsed laser $1 / \tau$. However, the typical pulse duration of a modelocked Ti:sapphire pulsed laser on the order of 100 fs - 1 ps provides sufficient bandwidth to address transitions between rotational and hyperfine states in a molecule. Pulses from both directions should arrive at the position of the ions at the same time, therefore the path length difference between two arms cannot exceed $ c \tau $ (about 30 $\mu$m for a 100 fs pulse).

To address the motion of the ions we detune the spectral beat note between the two laser beams from the two photon resonance  between molecular states $|m_1\rangle$ and  $|m_2\rangle$ by $\omega_t$, the frequency of a common motional mode. In this case the interaction Hamiltonian in the Lamb-Dicke limit has the form \cite{wineland}
\begin{equation}
\hat H_I = \hbar \eta  \Omega (a^{\dagger} \sigma_{-} + a \sigma_{+} )
\end{equation}
where $a^{+}$ ($a$) is the phonon creation (annihilation) operator for the common mode of motion of two ions, and $\sigma_{+} = |m_1\rangle\langle m_2|$ ($\sigma_{-} = |m_2\rangle\langle m_1|$) is the raising (lowering) operator for the transition. The change in the molecular quantum state is accompanied by a simultaneous change of the motional state of the ions. In the ground state of motion, the Rabi frequency of a sideband transition is equal to $\Omega_s = \eta \Omega$, where $\eta = k\sqrt{\hbar / 2 m \omega_t}$ is the Lamb-Dicke parameter, and  $k$ is the wave vector difference between the two laser beams.

The state of the molecular ion can be detected using quantum logic techniques \cite{rosenband,schmidt}. If the atomic and molecular ions were initially in a ground state of motion, detection of the quantum state of a molecular ion can be accomplished by driving a Raman transition that couples the state of the molecular ion  $|m_1\rangle$ to a collective motional mode of the ions and generates a phonon if the molecular ion was in the state $|m_1\rangle$. The phonon can be detected later by coupling the motion of the ions to the spin of an atomic ion, followed by the atomic ion state detection using the standard fluorescence technique.

Due to the periodic structure of the frequency comb, scanning the offset frequency $\nu_{AO}$ produces a spectrum that repeats itself every $f_{rep}$. Measurement of the absolute energy difference between two states therefore requires knowledge of both the offset frequency $\nu_{AO}$ and the comb index $M$. The latter can be determined by measuring the spectrum as a function of  the offset frequency $\nu_{AO}$ for several slightly different repetition rates and comparing the results \cite{junye_spectroscopy}.

\begin{figure}
\centering
\includegraphics[width=7cm]{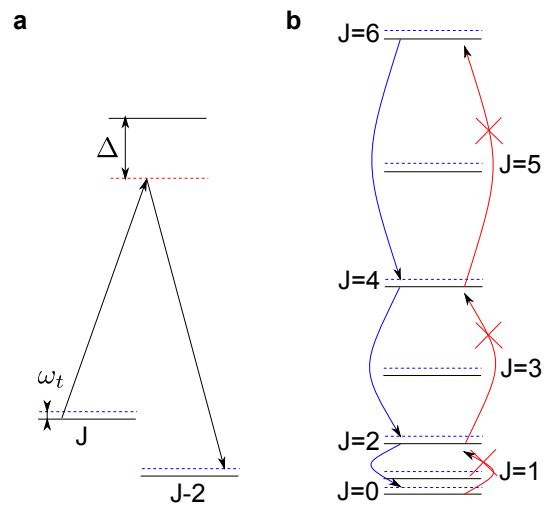}
\caption{Cooling of molecular rotational states using quantum logic. a) Stimulated Raman transition is driven on a 'blue" motional sideband so that Raman transition from rotational state $J$ to $J-2$ adds one phonon to the motional mode of the ions. b) If the molecule is cooled to the ground state of motion and phonons are continuously removed by sideband cooling of a co-trapped atomic ion, transitions that increase $J$ are not allowed, leading to an efficient build up of the population in the $J=0$ and $J=1$ states.  }
\label{fig:cooling}
\end{figure}

We can also prepare a molecular ion in the well defined quantum state by coupling internal and motional degrees of freedom. We start with atomic and molecular ions in the ground state of motion after sideband cooling and drive the $J, n=0 \rightarrow J-2, n=1$ Raman transition in the molecule, where $n$ is number of phonons in a given motional mode (see Fig. \ref{fig:cooling}). If the ions are in the ground state of motion, decrease of the angular momentum via the $J, n=0 \rightarrow J-2, n=1$ transition is allowed. However, the opposite transition $J -2 \rightarrow J$  is not allowed, since it requires subtracting motional energy from the ions that are already in the ground state.  If after each step phonons are removed from the trap by sideband cooling, the angular momentum $J$ of the molecule decreases, leading to a buildup of population in the $J=0$ or $J=1$ state. Any cooling requires some  dissipation process. In our scheme, the coupling between internal and motional degrees of freedom of the molecule is coherent, and dissipation is achieved via cooling of the common motion of the molecular and atomic ions.

It is interesting to note that this cooling scheme  can be made more efficient if we use the periodic structure of a frequency comb. For example, the proper choice of repetition rate and offset frequency allows us to drive transitions between at least two pairs of levels at the same time. In some special cases, for example, heavy molecules with no hyperfine structure confined in a tight trap, when the centrifugal distortion constant $D$ is small compared to the rotational constant $B$, the splitting between energy levels $E(J+2) - E(J) \simeq 2 B (3 + 2 J) (1 + \frac{2 D} { B} (3 + 3 J + J^2))$ increases almost linearly with $J$ and can be matched to the periodic structure of a frequency comb, which allows even more pairs of levels to be addressed  simultaneously.

The number of Zeeman sublevels in the $J+2$  state is larger than in the state  $J$, therefore it is impossible to couple all the sublevels of the upper state to the lower one simultaneously via Raman transition, which leads to population trapping in the states with higher angular momentum. To avoid it we can alternate the polarization of the Raman beams between several configurations to make sure that all sub-levels  in the upper $J+2$ state are coupled to the sublevels of  the lower $J$ state. Alternatively, we can apply a weak magnetic field to mix the Zeeman sublevels. However, the latter approach is less desirable since Zeeman splitting can increase the number of energy levels one has to address. 

As a possible first step towards the implementation of this scheme, we consider a SiO$^+$ molecular ion trapped together with an  Yb$^{+}$ atomic ion. The spectral properties of SiO$^{+}$ are known \cite{francois}. The wavelength for the $X^{2}\Sigma^{+}(\nu = 0) \rightarrow A^{2} \Pi (\nu=0)$ transition is about 414 nm, and the $X^{2}\Sigma^{+}(\nu=0) \rightarrow B^{2} \Sigma^{+}(\nu=0)$ transition wavelength is near 383 nm, close to the 370 nm transition of atomic Yb$^{+}$ ion. Nearly diagonal Franck-Condon factors for the $X^{2} \Sigma^{+}(\nu=0) \rightarrow B^{2} \Sigma^{+}(\nu=0)$ transition in SiO$^+$ maximize the two-photon Rabi frequency. The absence of  hyperfine structure for the most abundant molecule Si$_{28}$O$_{16}$ simplifies the energy level structure.

The efficiency of sympathetic cooling depends on the mass ratio between the two ions. A 1:1 mass ratio provides the highest efficiency due to the largest momentum transfer. However, sympathetic cooling was demonstrated for a wide range of masses, for example for Al$^+$ /Be$^+$ system with a 3:1 mass ratio \cite{schmidt}, and can also be feasible in the Yb$^{+}$ and SiO$^{+}$ system where the ratio is 4:1. 

The energy of rotational states in the ground electronic state $X$ and vibrational state $\nu=0$ is $E_{X}(J) = B J (J + 1) + D J^2 (J+1)^2$, where for SiO$^+$ $B = 21.51$ GHz and $D=33.1$ kHz \cite{francois}. At room temperature about $98\%$ of population is distributed among states with angular momentum from $J=0$ to $J=35$ of the lowest vibrational state $\nu = 0$. While it is still possible to apply the cooling scheme described in the paper directly, the large number of populated levels and magnetic fields can significantly increase the number of required cooling steps. Therefore some initial cooling of the rotational degrees of freedom is desirable.

\begin{figure}
\centering
\includegraphics[width=8.5cm]{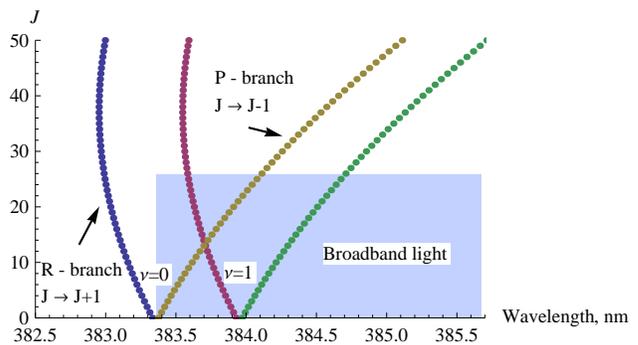}
\caption{Optical pumping scheme for SiO$^+$ molecule. The P ($J \rightarrow J-1$ transitions), and R-branched ($J \rightarrow J+1$ for the SiO$^+$ molecule are well separated. Number of populated rotational states in a molecule can be reduced using optical pumping with spectrally filtered broadband light that excites $J \rightarrow J-1$, but not  $J \rightarrow J+1$).
}
\label{fig:pumping}
\end{figure}

One approach to decrease the number of populated rotational states is optical pumping with spectrally shaped broadband light \cite{viteau,odom}. For the $X^{2} \Sigma^{+}(\nu=0) \rightarrow B^{2} \Sigma^{+}(\nu=0)$ transition in SiO$^+$, the $R$ ($J \rightarrow J+1$ transitions) and $P$ ($J \rightarrow J-1$ transitions) branches are well separated (see Fig. \ref{fig:pumping}). Light with the intensity $dI / d \lambda = 1 $mW/nm can be produced by a high power UV LED, and spectral filtering with the resolution better than 0.2 nm is easily achievable with a diffraction grating. Focusing this light into a 50 $\mu$m spot provides a scattering rate of about $10^5$ photons per second. This scattering rate can decrease the number of populated rotational states to less than 10 in about a millisecond for final preparation of states with the quantum logic scheme.

For a realistic average intensity of the modelocked pulsed laser at the ion position of about 1000 W/cm$^2$,  laser detuning $\Delta / 2 \pi = 20$ THz, lifetime of $B^{2}\Sigma^{+}$ excited state $1 / \gamma = 70$ ns, and  the saturation intensity of molecular transition $I_{sat} = 45$ W / m$^2$  one can expect time averaged Rabi frequency (\ref{eq:omega}) on the carrier transition to be $\Omega_0 / 2 \pi = 0.2 $ MHz. For a Lamb-Dicke parameter $\eta = 0.1$, the duration of the $\pi$ pulse on a sideband transition of a molecule is about 100 $\mu$s. The sideband cooling step should have a comparable duration. Assuming that the rotational population is distributed among 10 lowest $J$ states, and that it takes about 10 cycles to bring the population from $J$ to $J-2$ level, we estimate the total cooling time to be on the order of 20 ms, much faster than the $\sim20$ s achievable by alternative schemes that rely on  blackbody radiation for repumping between rotational states \cite{drewsen,schiller}.

Spontaneous emission will likely remove the molecule out of the cooling cycle and leave it in a different  rotational, or even different vibrational state. Assuming that the bandwidth of the mode locked pulsed laser $\sim 1/\tau << \Delta$ the rate of spontaneous emission can be estimated as $R_s = 2 \gamma \Omega / \Delta = 0.3 $ s$^{-1}$, much slower than the duration of the cooling sequence.

The quantum logic scheme presented above is general, it does not impose restrictive requirements on molecular structure, and can be applied to a wide range of molecular ions. Experimental realization of this scheme can open the way to control quantum states of molecular ions and find application in precision measurements, quantum information, and quantum chemistry.

Authors would like to thank Brenda Chng, Wes Campbell, Alex Kuzmich, Chris Monroe, Dave Wineland and Jun Ye for valuable discussions, and to Dietrich Leibfried for providing a manuscript \cite{didi} before publication. The work is supported by the National Research Foundation and the Ministry of Education, Singapore.

During the preparation of the manuscript we became aware of similar work by Dietrich Leibfried \cite{didi}.


\begin{thebibliography}{99}
\bibitem{junye1}  Carr, L. D.,  DeMille, D., Krems, R. V.,  \& Ye, J. Cold and ultracold molecules: science, technology and applications { \it New J. Phys.} \textbf{11}, 055049 (2009)
\bibitem{junye3}  Hudson, E. R., Lewandowski, H. J., Sawyer, B. C.,  \& Ye J. Cold molecule spectroscopy for constraining the evolution of the fine structure constant. {\it Phys. Rev. Lett.} \textbf{96} 143004 (2006).
\bibitem{kozlov2} Chin, C., Flambaum, V. V.,  \& Kozlov, M. G. Ultracold molecules: new probes on the variation of fundamental constants. {\it New J. Phys.} \textbf{11}, 055048 (2009)
\bibitem{willeke} Gottselig, M., Quack, M., Stohner, J.,  \& Willeke, M. Mode-selective stereomutation tunneling and parity violation in HOClH$^+$ and H$_2$Te$_2$.  {\it Int. J. Mass Spec.} \textbf{233}, 373-384 (2004).
\bibitem{kozlov3} DeMille, D.,  Kahn, S. B., Murkphree, D., Rakhmlow, D. A. \& Kozlov, M. G. Using molecules to measure nuclear spin-dependent parity violation. {\it Phys. Rev. Lett.} \textbf{100}, 023003 (2008).
\bibitem{cornell} Stutz, R.P. \& Cornell, E.A. Search for the electron EDM using trapped molecular ions. {\it Bull. Am. Soc. Phys.} \textbf{89}, 76 (2004).
\bibitem{softley} Willitsch, S., Bell, M. T., Gingell, A. D. \& Softley, T. P. Chemical applications of laser- and sympathetically-cooled ions in ion traps. {\it Phys. Chem. Chem. Phys.} \textbf{10}, 7200-7210 (2008).
\bibitem{andre}  Andre, A. {\it et al.} A coherent all-electrical interface between polar molecules and mesoscopic superconducting resonators. {\it Nature Physics} \textbf{2}, 636-642 (2006).
\bibitem{rosenband} Rosenband, T. {\it et al} Frequency Ratio of Al$^+$ and Hg$^+$ Single-Ion Optical Clocks; Metrology at the 17th Decimal Place. {\it Science}  \textbf{319}, 1808-1812 (2008).
\bibitem{schmidt} Schmidt, P. O., Rosenband, T., Langer, C., Itano, W. M., Bergquist, J. C. \& Wineland, D. J. Spectroscopy Using Quantum Logic. {\it Science} \textbf{309}, 749-752 (2005).
\bibitem{rempd} Van Heijnsbergen, D., Jaeger, T. D., von Helden, G., Meijerand, G. \& Duncan, M. A. Infrared Spectroscopy of Al+-(benzene) in the Gas Phase. {\it Chem. Phys. Lett.} \textbf{364}, 345-351 (2002).
\bibitem{drewsen} Staanum, P. F. , Hojbjerre, K., Skyt, P. S. , Hansen, A. K. \& Drewsen, M. Rotational laser cooling of vibrationally and translationally cold molecular ions. {\it Nature Physics} \textbf{6}, 271-274 (2010).
\bibitem{schiller} Schneider, T., Roth, B., Duncker, H., Ernsting, I. \& Schiller, S. All-optical preparation of molecular ions in the rovibrational ground state. {\it Nature Physics} \textbf{6}, 275-278 (2010).
\bibitem{vogelius}  Vogelius, I. S.,  Madsen, L. B. \& Drewsen, M. Rotational cooling of molecular ions through laser-induced coupling to the collective modes of a two-ion Coulomb crystal  {\it J. Phys. B:  At. Mol. Opt. Phys.} \textbf{39}, S1267-S1280 (2006).
\bibitem{willitsch} Tong, X., Winney, A. H. \& Willitsch, S. Sympathetic cooling of molecular ions in selected rotational and vibrational states produced by threshold photoionization.  {\it Phys. Rev. Lett.} \textbf{105}, 143001 (2010).
\bibitem{viteau} Viteau, M., {\it et. al.}, Optical Pumping and Vibrational Cooling of Molecules {\it Science} \textbf{321}, 232-234 (2008).
\bibitem{odom} Lien, C.-Y., Williams, S. R. \& Odom, B. Optical pulse-shaping for internal cooling of molecules. {\it arxiv:1104.3177}.
\bibitem{hudson} Hudson, E. R., Method for producing ultracold molecular ions. {\it Phys. Rev. A} \textbf{79}, 032716 (2009).
\bibitem{morigi} Morigi, G., Pinske, P.W.H., Kowalewski, M. \& de Vivie-Riedle, R. Cavity cooling of internal molecular motion. {\it Phys. Rev. Lett.} \textbf{99}, 073001 (2007).
\bibitem{lev} B. L. Lev {\it et al.}, Prospects for the cavity-assisted laser cooling of molecules. {\it Phys. Rev. A} \textbf{77}, 023402 (2008).
\bibitem{lazarou} Lazarou, C., Keller, M. \& Garraway, B. M. Molecular heat pump for rotational states. {\it Phys. Rev. A} \textbf{81}, 013418 (2010).
\bibitem{vogelius2} Vogelius, I. S., Madsen, L. B. \& Drewsen, M. Probabilistic state preparation of a single molecular ion by projection measurement. {\it J. Phys. B: At. Mol. Opt. Phys. B} \textbf{39}, S1259-S1266 (2006).
\bibitem{hayes} Hayes, D., {\it et al.}  Entanglement of Atomic Qubits Using an Optical Frequency Comb. {\it Phys. Rev. Lett.} \textbf{104}, 140501 (2010).
\bibitem{junye4} Shapiro, E. A.,  Pe'er, A., Ye, J. \& Shapiro, M. Piecewise Adiabatic Population Transfer in a Molecule via a Wave Packet, {\it Phys. Rev. Lett.} \textbf{101}, 023601 (2008).
\bibitem{damop} Ding, S. \& Matsukevich, D. Control and manipulation of cold molecular ions. {\it Bull. Am. Phys. Soc.}, \textbf{56}, No 5, L1.00015 (DAMOP2011).
\bibitem{monroe} Monroe, C., {\it et. al} Resolved-Sideband Raman Cooling of a Bound Atom to the 3D Zero-Point Energy. {\it Phys. Rev. Lett.} \textbf{75}, 4011-4014  (1995).
\bibitem{wineland} Wineland, D. J.,  {\it et al.} Experimental Issues in Coherent Quantum-State Manipulation of Trapped Atomic Ions. {\it Journal of Research of the National Institute of Standards and Technology} \textbf{103}, 259 (1998).
\bibitem{junye_spectroscopy}  Stowe, M. C., Thorpe, M. J., Pe'er A., Ye, J.,  Stalnaker, J. E., Gerginov, V. \& Diddams, S. A. Direct frequency comb spectroscopy.
{\it Advances in Atomic, Molecular and Optical Physics}, \textbf{55}, 1-60 (2008).
\bibitem{didi} Leibfried D. Quantum state preparation and control of single molecular ions. {\it arXiv:1109.0208}.
\bibitem{francois} Cai, Z.-L. \& Francois, J. P. Ab initio study of the electronic spectrum of the SiO$^+$ cation. {\it J. Mol. Spec.} \textbf{197}, 12-18 (1999).
\end{thebibliography}
\end{document}